\documentclass[aps,prl,twocolumn,superscriptaddress]{revtex4-1}
\usepackage{amssymb}
\usepackage{physics}
\usepackage{graphicx}
\usepackage{amsmath}
\usepackage{amsthm}
\usepackage{amsfonts}
\usepackage[T1]{fontenc}
\usepackage{array}
\usepackage{multirow}
\usepackage{color}
\usepackage{esint}
\usepackage{bm}
\usepackage{color}
\usepackage{bbm}
\usepackage{hyperref}
\usepackage{babel}
\usepackage{titlesec}
\usepackage{MnSymbol}
\newcommand{\kket}[1]{\left\lVert #1 \right\rrangle}
\newcommand{\bbra}[1]{\left \llangle #1 \right\rVert}
\newcommand{\mean}[1]{\langle #1 \rangle}
\newcommand{\cl}{\mathcal{L}}

\begin{document}
	\title{Dissipative prethermal discrete time crystal}
	
	\author{DinhDuy Vu}
	\author{Sankar Das Sarma}
	\affiliation{Condensed Matter Theory Center and Joint Quantum Institute, Department of Physics, University of Maryland, College Park, Maryland 20742, USA}
	
	\begin{abstract}
    An ergodic system subjected to an external periodic drive will be generically heated to infinite temperature. However, if the applied frequency is larger than the typical energy scale of the local Hamiltonian, this heating stops during a prethermal period that extends exponentially with the frequency. During this prethermal period, the system may manifest an emergent symmetry that, if spontaneously broken, will produce sub-harmonic oscillation of the discrete time crystal (DTC). We study the role of dissipation on the survival time of the prethermal DTC. On one hand, a bath coupling increases the prethermal period by slowing down the accumulation of errors that eventually destroy prethermalization. On the other hand, the spontaneous symmetry breaking is destabilized by interaction with environment. 
   The result of this competition is a non-monotonic variation, i.e. the survival time of the prethermal DTC first increases and then decreases as the environment coupling gets stronger.
	\end{abstract}

    \maketitle
    \textit{Introduction - } For static systems, the spontaneous symmetry breaking (SSB) is paradigm dividing matter into phases, most notably through the Landau-Ginzburg theory. It is important to ask whether SSB can manifest in dynamic systems, in particular those with time translation symmetry.The no-go theorem prohibits the SSB of continuous time translation symmetry \cite{Bruno2013,Watanabe2015}.  Many attempts have been made to circumvent this situation but are still debatable using long-range multi-spin interaction \cite{Kozin2019,Khemani2020c,Kozin2020} or interacting gauge theory \cite{Ohberg2019,Syrwid2020,Ohbergreply,Syrwid2020b,Syrwid2021}. On the other hand, SSB is established to manifest under discrete time translation symmetry (in Floquet systems) \cite{Khemani2016,Else2016,Moessner2017a,Yao2017}, producing an exotic phase called the discrete time crystal (DTC). The signature of this phase is the many-body collective response exhibiting a longer periodicity than that of the drive, usually an integer multiple. Several experiments have successfully created DTC in atomic systems or quantum simulators \cite{Zhang2017,Choi2017,Ippoliti2021,Mi2022,Frey2022,Kyprianidis2021,Beatrez2021,Beatrez2022}. 
    
    One of the main problem in realizing DTC is the heating to infinite temperature by the periodic driving field. Therefore, it is crucial to prevent thermalization or at least delay it by a sufficiently long time. The first discovered strategy for this task is to utilize many-body localization (MBL) by introducing disorder to the Hamiltonian \cite{Khemani2016,Else2016,Moessner2017a} or a static electric field \cite{Bhakuni2020}. Since MBL violates the Eigenstate Thermalization Hypothesis, the information of the initial state still persists in the long-time dynamics \cite{Nandkishore2015,Abanin2019}, thus we can expect the DTC, if protected by MBL, to survive to infinite time. However, disorder-induced MBL requires tuning and might be difficult to engineer, not to mention that the validity of MBL in the thermodynamic limit is still not settled \cite{Chaos2020,Sierant2022,Tu2022}. Another approach - prethermalization, on the other hand, only requires the applied frequency to be larger than the smallest energy scale of the Hamiltonian \cite{Abanin2017b, Abanin2017,Else2017,Zeng2017,Machado2020,Natsheh2021}.  During the prethermal regime, the dynamics manifests an emergent symmetry with exponentially small error even though it is not an exact symmetry of the Hamiltonian. If the symmetry is represented by a $\mathbb{Z}_N$ group and is spontaneously broken,  the sub-harmonic response will emerge in some collective degree of freedom with periodicity $NT$. In some exotic case, the multiplicity can even be non-integer \cite{Pizzi2021}.  Unlike the MBL proposal, the prethermal DTC survives only a finite time before being eventually thermalized, but this time can be exponentially extended by simply increasing the driving frequency. This characteristic has been observed \cite{Kyprianidis2021,Beatrez2021,Beatrez2022}. The key point is that prethermal DTC should exist independent of whether MBL exists or not.
    
The above strategies were initially developed for closed systems which may not reflect the realistic situation as a system is always coupled to the environment not only in terms of heat bath but also noise. In the presence of a bath, MBL is most likely destroyed, while the fate of prethermal DTC is less straightforward. For example, a cold bath can potentially preserve the DTC to infinite time as it absorbs excessive heat generated by the drive \cite{Else2017}. We show that the prethermal formalism extends to open systems, i.e, up to a time scale exponentially long in the applied frequency, the dynamics is approximated by that under a time-independent Lindbladian. However, unlike a constant Hamiltonian that defines a conserved energy, a constant Lindbladian eventually drives the system to a steady state regardless of the initial condition. Therefore, the effect of disspation on the observability of time crystal is more complicated, involving different time scales.
    
\begin{figure}
     \centering
     \includegraphics[width=0.48\textwidth]{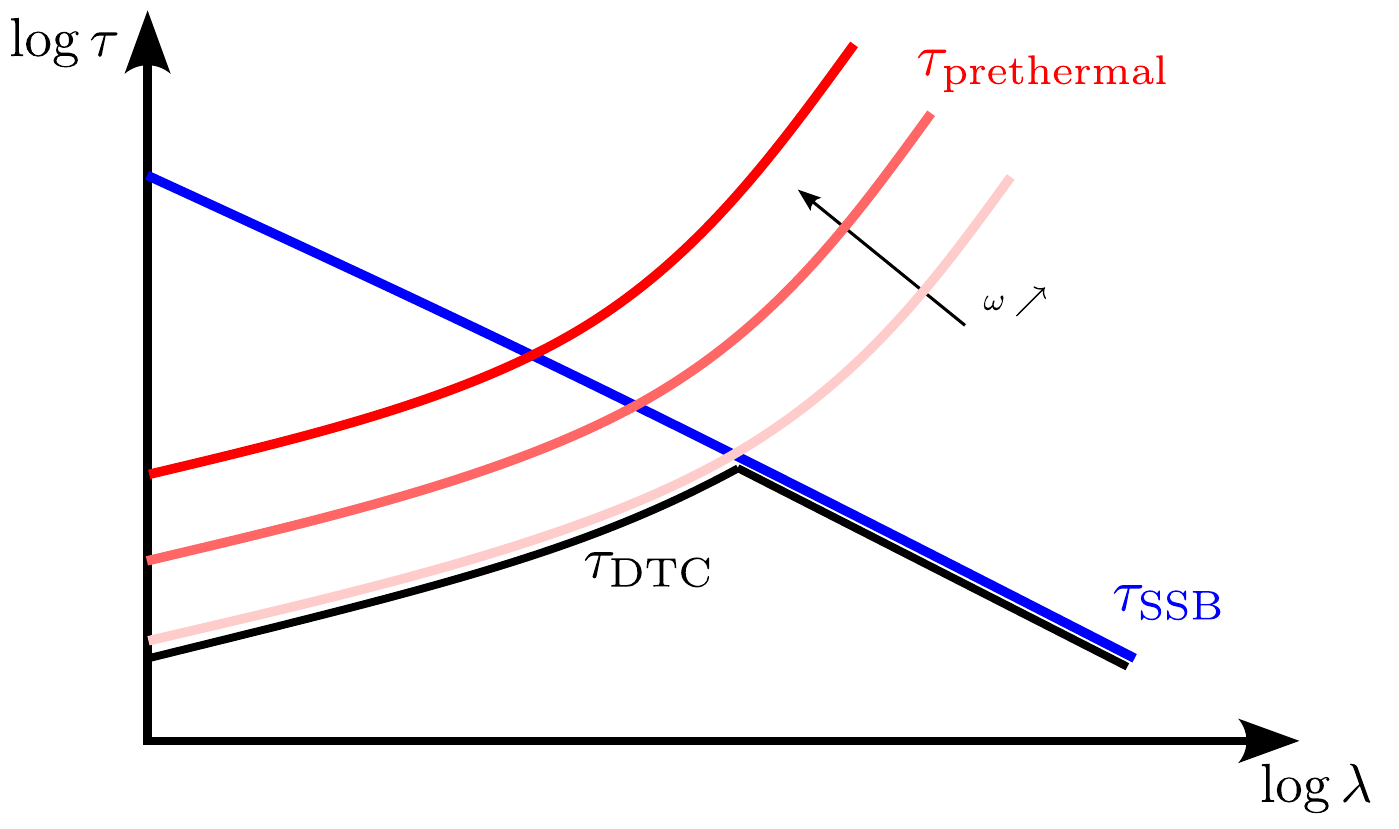}
     \caption{Variation of the time window of the prethermalization, quasi-SSB and DTC with respect to the dephasing noise strength. The survival time of DTC is the shorter one between the prethermal and the SSB lifetime. \label{fig1}}
    \end{figure}

    The emergence of prethermal DTC is based on two conditions: the prethermal regime and the SSB of the emergent symmetry. We note that a true SSB has infinite lifetime by definition, but in a finite system, we can only achieve quasi-SSB whose large fluctuations of the order parameter must eventually vanish to recover the symmetry. The shorter lifetime between the prethermalization and quasi-SSB thus determines the survival time of DTC.  Our result is summarized in Fig.~\ref{fig1}. We show that as the environment coupling gets stronger, the prethermal regime is extended, while the quasi-SSB lifetime is reduced, leading to a non-monotonic behavior of the DTC. Additionally, in the early increasing phase, the exponential dependence on the applied frequency is prominent; while in the later decreasing phase, the frequency becomes irrelevant. This statement is demonstrated numerically in the main text and argued analytically in the Supplemental Material \cite{Supp}. We mention that for some specific forms of noise operators, the decreasing trend may start at a very weak noise amplitude, making the first increasing phase almost invisible. Before proceeding to the details, we contrast our study with the disspative time crystal that does not require emergent symmetry or MBL \cite{Kessler2019,Gambetta2019,Wang2020,Sakurai2021,Phatthamon2022}. This proposal, on the other hand, requires the steady states to be degenerate or quasi-degenerate, which does not hold generically but can be enforced by tuning the system across a phase transition. As such, DTC only emerges near the critical point of a phase transition and is protected by the dissipative gap between the quasi-degenerate manifold and the rest of the spectrum, necessitating a finetuned engineering of the Lindbladian. 
    
    \textit{Prethermalization in open systems - }We first translate the derivation of prethermal DTC from the unitary evolution in closedd systems to the Liouvillian evolution of open systems. The dynamics is driven by a time-periodic Liouvillian $\mathcal{L}(t)=\mathcal{L}(t+T)$ with $T=2\pi/\omega$ being a fixed applied period. We assume that the Liouvillian contains both unitary and dissipative parts and is described by the Lindblad equation
    \begin{equation}\label{eq:Lindblad}
    	\cl[\rho] = -i[H,\rho] + \sum_j \lambda_j \left( L_j\rho L_j^\dagger -\frac{1}{2}\{L_j^\dagger L_j,\rho\} \right)
    \end{equation}
    In our work, the environmental coupling manifests as local dephasing noise so the channel index $j$ in the dissipative part is also the site index and we choose $\lambda_j =\lambda$. Another approach to access open system is to solve the stochastic evolution of a pure state, known as Heisenberg-Langevin equation \cite{Tuquero2022}.
    
    To make connection with the standard unitary evolution used to derive prethermalization, we promote the density matrix to a supervector $\kket{\rho}$ ($4^L$-vector) and the Liouvillian $\cl$ to a superoperator $\hat{\cl}$ ($4^L\times4^L$-matrix). As a result, the density matrix at a time $t$ is given by an evolution similar to the unitary case $	\kket{\rho(t+T)}=\hat{U}_f\kket{\rho(t)}=\mathcal{T} e^{\int_0^T\hat{\cl}(s)ds}\kket{\rho(0)}$ with $\mathcal{T}$ being the time ordering operator.
   The expectation value of an operator is also brought into a Schr\"{o}dinger-like form $\mean{O(t)} = \bbra{\mathbbm{1}} O \kket{\rho(t)}$ with $\kket{\mathbbm{1}}$ corresponding to the identity operator. Here, we use the normal definition of vector inner product. The derivation of the slow heating is similar to the unitary case except for $\hat{\cl}$ being non-Hermitian. Due to the assumption that the dissipative dynamics is much weaker than the coherent one, the emergent symmetry is given by $X=\mathcal{T} e^{-i\int_0^T H_0(s)ds}$ satisfying $X^N=\mathbbm{1}$ so that in the DTC phase, the system repeats itself after $N$ cycles. Here, $H_0(t)$ is the dominant part of the driving Hamiltonian, in particular the energy scale $J_{\text{res}}$ of the residue $H-H_0$ is much less than $1/T$. By applying similar iterative optimization as for closedd systems \cite{Abanin2017,Else2017,Machado2020}, we arrive 
    \begin{equation}
       e^{\hat{A}}\hat{U}_fe^{-\hat{A}} = \hat{X} \mathcal{T} e^{\int_0^T [\hat{D} + \hat{V}(s)]ds} 
    \end{equation}
    where $e^{\hat{A}}$ is a time-independent trace-preserving map, $[\hat{X},\hat{D}]=0$, and $\hat{X}$ is the superoperator promotion of $X[.]X^\dagger$. Importantly, the norm of residue term $\hat{V}$ is exponentially suppressed by the driving frequency. Because $\hat{D}$ is a complete positive trace preserving (CPT) map generator and has a much larger norm than the other terms, the ``rotated'' $e^{\hat{A}}\hat{U}_fe^{-\hat{A}}$ is also a CPT map. Within a time period $\sim e^{\omega/J_\text{res}}$ only $\hat{D}$ is relevant to the dynamics and $\hat{X}$ becomes the emergent symmetry of the system. If $\hat{X}$ is spontaneously broken, the prethermal DTC is observable. Up to this point, it appears that DTC in open systems must behave similar to its closedd system counterpart. In the following, we point out two key distinctions in the residual error and the stability of SSB.
    
    Approximating the exact Lindbladian by that only the symmetric time-independent $\hat{D}$ incurs errors which accumulate with time and eventually suppress the prethermal period, thermalizing the system. For a local observable $O$, the accumulation rate is bounded by the Lieb-Robinson velocity and whether the interaction is short or long-ranged \cite{Sweke2019,Guo2021}.Obviously, decreasing Lieb-Robinson velocity necessarily increases $\tau_{\text{prethermal}}$. However, since our noise model is strictly onsite, it is not obvious how $\lambda$ can enter the expression of the Lieb-Robinson velocity $v$ which is responsible for the inter-site information propagation, not to mention that the noise magnitude is seemingly irrelevant compared to the other scales. On the other hand, if the bath coupling results in a unique steady state then information of the initial state must be lost eventually. From this long-time limit, Ref.~\cite{Descamps2013} suggests that the velocity may decay exponentially in the presence of environment coupling. In our case, the onsite noise induces a deceleration rate so that $\partial_t v(t) \propto -\lambda$ As a result, we can relate the dissipative prethermal period with that in the dissipationless limit
    \begin{equation}\label{tprethermal}
    	\tau_{\text{prethermal}}^\lambda = \tau_{\text{prethermal}}^0\left( 1-C \lambda \tau_{\text{prethermal}}^0 \right)^{-1}
    \end{equation}
    with $C$ being an $\mathcal{O}(1)$ constant \cite{Supp}. Therefore, even though $\lambda$ is the smallest scale by our assumption, it can still result in visible effects on the DTC lifetime when accompanying the exponentially long $\tau^0_{\text{prethermal}}$.
    
    The second aspect where the dissipative nature of the system becomes relevant is the survival time of the quasi-SSB. In a closedd system, energy is conserved so a single excitation, e.g. spin flip, is not favorable. On the other hand, the creation of multiple excitations so that the ground state is mapped to its degenerate partner requires higher orders of perturbation and hence is suppressed exponentially. Therefore, even for finite system where the ground state must be symmetric, the energy splitting can be exponentially small. Unsurprisingly, the lifetime of the quasi-SSB scales exponentially with system size and is usually taken to be infinity even for systems with moderate size. In an open system, however, energy can be exchanged with the environment to stabilize the excitation, thus destabilizing the quasi-SSB. In fact, the finite-size effect is much more severe in open systems through the fact that $\tau_{\text{SSB}} \sim L^{1/2}$ in open 1D chains \cite{Wilming2005}. This necessitates a careful analysis on the survival time of the quasi-SSB. In the Supplemental Material \cite{Supp}, we show that the decay rate of the quasi-SSB increases monotonically with the bath coupling strength. Unlike the effect of noise on the prethermalization, different operational forms of noise $L_j$ may lead to vastly different decay rates. Under some form of dephasing noise, the decreasing slope (see Fig.~\ref{fig1}) becomes much more prominent so that the increasing slope is most likely unobservable.
    
    \textit{Numerical model - }As a demonstration, we study the driven Heisenberg chain subjected to dephasing noise. Referring to Eq.~\eqref{eq:Lindblad}
    \begin{equation}
    	H(t) = \sum_i \frac{\boldsymbol{h}(t)}{2}\boldsymbol{\sigma}^i + \frac{J_{xx}}{4}\sigma_x^i\sigma_x^{i+1} + \sum_{j>i} \frac{J}{4|j-i|^\alpha}\sigma_z^i\sigma_z^j
    \end{equation} 
    where $\boldsymbol{\sigma}^i=\{\sigma_x^i,\sigma_y^i,\sigma_z^i\}$ is the collection of Pauli matrices at site $i$. The periodic Zeemann field $\boldsymbol{h}$ is given by
    \begin{equation}
    	\boldsymbol{h}(t)\boldsymbol{\sigma}^i = \begin{cases}
    		\boldsymbol{h}_s\boldsymbol{\sigma}^i & \text{ for } nT < t \le (n+1)T - t_p \\
    		\pi t_p^{-1} \sigma_x^i & \text{ for } (n+1)T -t_p < t \le (n+1)T 
    	\end{cases}
    \end{equation}
    with finite but small constant $\boldsymbol{h}_s$ and in the limit $t_p\to 0$. The former condition ensure the drive frequency is larger than other energy scales and the latter limit describes an instant spin flip. We note that the physics does not change significantly given a finite-width $\pi-$pulse. With this set up, the emerging symmetry arises from the dominant $\pi-$pulse sequence and is given by $X=\prod_j \sigma_x^j$. Since $X^2=\mathbbm{1}$, the DTC oscillation features doubled periodicity of $2T$. The long-range $zz$ interaction with $1\le \alpha \le 2$ is vital to drive the transition to the spontaneous Ising symmetry breaking phase \cite{Dyson1969,Bhakuni2021}. Lastly, the term $J_{xx}$ breaks integrability so that the symmetrized $\hat{D}$ is not trivially diagonal. 
    
    For the bath coupling, we use two representative cases of dephasing noise: (i) $L_j=s_z^j$ and (ii) $L_j=s_x^j$ where the quantum channel index $j$ is also the site index. Physically, these quantum channel describes the coupling between the spin and an isolated harmonic oscillator reservoir sitting on the same site. In their respective basis, the dephasing noise keeps the diagonal entries of the density matrix while suppressing off-diagonal ones. In both cases, we set $\lambda_j=\lambda$ thus the system does not have any spatial disorder. Because the entanglement grows much faster under long-range interaction, we limit the system to $L=12$ and use exact diagonalization to evaluate the long-time behavior \cite{Supp}. We check that a closed analog of this rather small lattice is already sufficient to demonstrate all the signatures of the DTC.
    
    \begin{figure}
    \centering
    \includegraphics[width=0.48\textwidth]{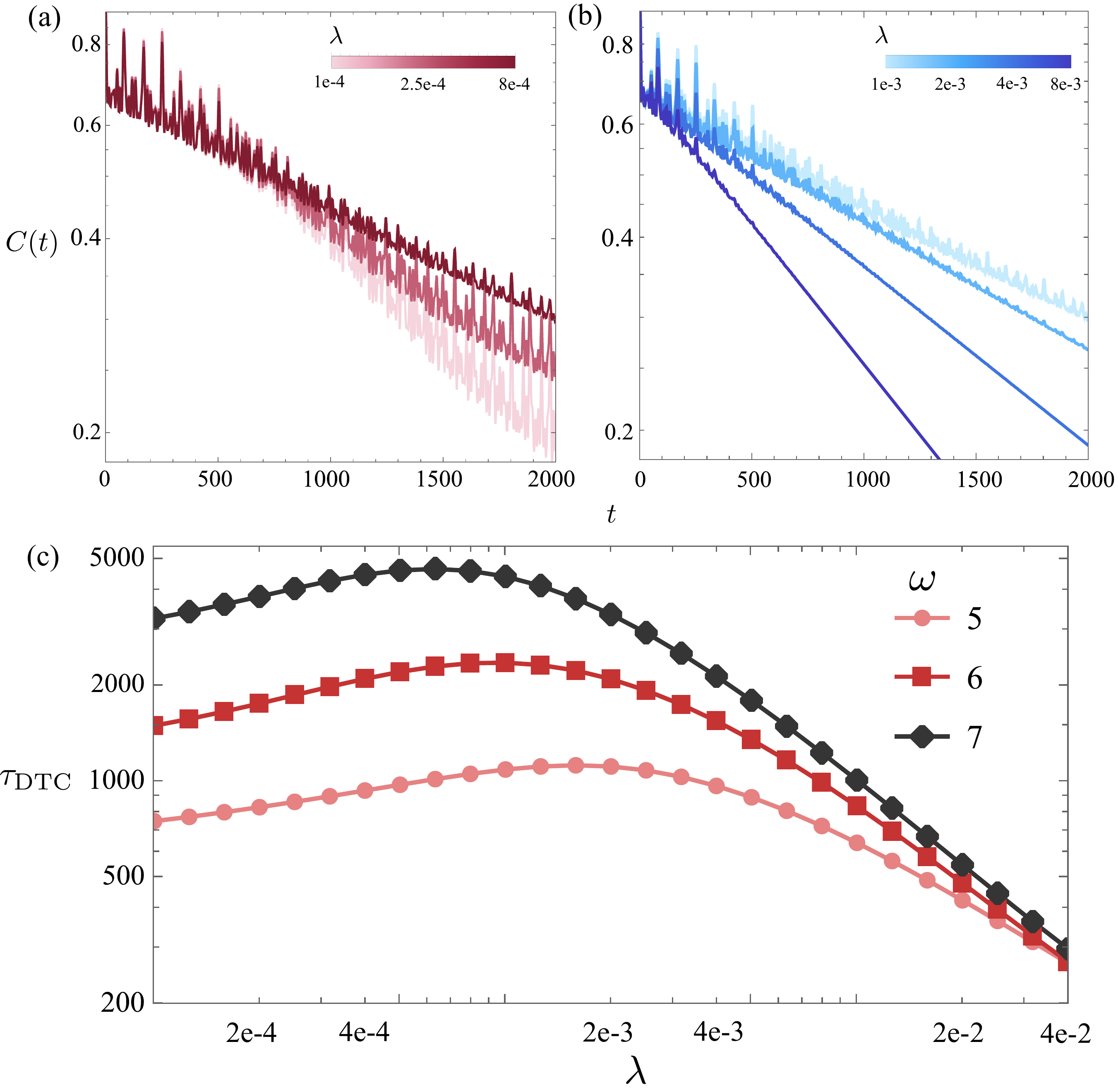}
    \caption{$2T$-DTC oscillation sampled at $t=2kT$ at $\omega=6$ within the gain range (a) and the loss range (b). The color bar is logarithmic in the dephasing strength. (c) Lifetime of the time crystal for different frequency. In the gain regime, the DTC survival time is enhanced exponentially by increasing the applied frequency; while in the loss regime, the applied frequency is almost irrelevant. \label{fig2}}
    \end{figure}
    
    \textit{$S_z-$dephasing noise - }With the $\mathbb{Z}_2$ Ising symmetry as the emergent symmetry, an observable associated with an extensive degree of freedom in the DTC phase should repeat itself after $2T$. A typical choice is the magnetization or magnetization density. In this work, we compute the normalized magnetization, following Ref.~\cite{Machado2020}
    \begin{equation}
    	C(t)=L^{-1}\sum_j \mean{\sigma_z^j(t)}\mean{\sigma_z^j(0)}.
    \end{equation} 
    By definition, $C(0)$ is always normalized to unity. In Fig.~\ref{fig2}(a), we show the stroboscopic $C(t)$ at even cycles ($C(t)$ at odd cycles is the reflection through zero) for weak dephasing noise. Here, we can see that initially the survival time of the DTC increases with the bath coupling strength. It is remarkable that this extension is significant even when $\lambda$ is much smaller than any energy scales of the Hamiltonian. This is the result of Eq.~\ref{tprethermal}, specifically $\lambda\sim \mathcal{O}(10^{-3})$ but when accompanied by $\tau_{\text{prethermal}^0} \sim \mathcal{O}(10^3)$ can yield visible effect on the DTC lifetime. As we further increase the noise amplitude, as shown in Fig.~\ref{fig2}(b), the DTC survival time begins to decrease after $\lambda=0.001$. 
    
    To understand these contrasting behaviors, we extract the lifetime $\tau_{\text{DTC}}$ by fitting $C(t)$ to an exponential decay and compare among different drive frequencies increases from 5 to 6 and 7, as shown in Fig.~\ref{fig2}(c). It is clear that in the increasing phase, the prethermal protection is apparent, i.e., $\tau_{\text{DTC}}$ scales exponentially with $\omega$.  Unlike the previous increasing phase, in this phase, there is no protection by the drive frequency, showing that $\tau_{\text{DTC}}$ is now bounded by a different time scale - the survival time of the quasi-SSB. We also emphases the dephasing strength where the DTC lifetime peaks shifts toward lower $\lambda$ as $\omega$ increases, consistent with the picture we described in Fig.~\ref{fig1}. 
    
\begin{figure}
	\centering
	\includegraphics[width=0.5\textwidth]{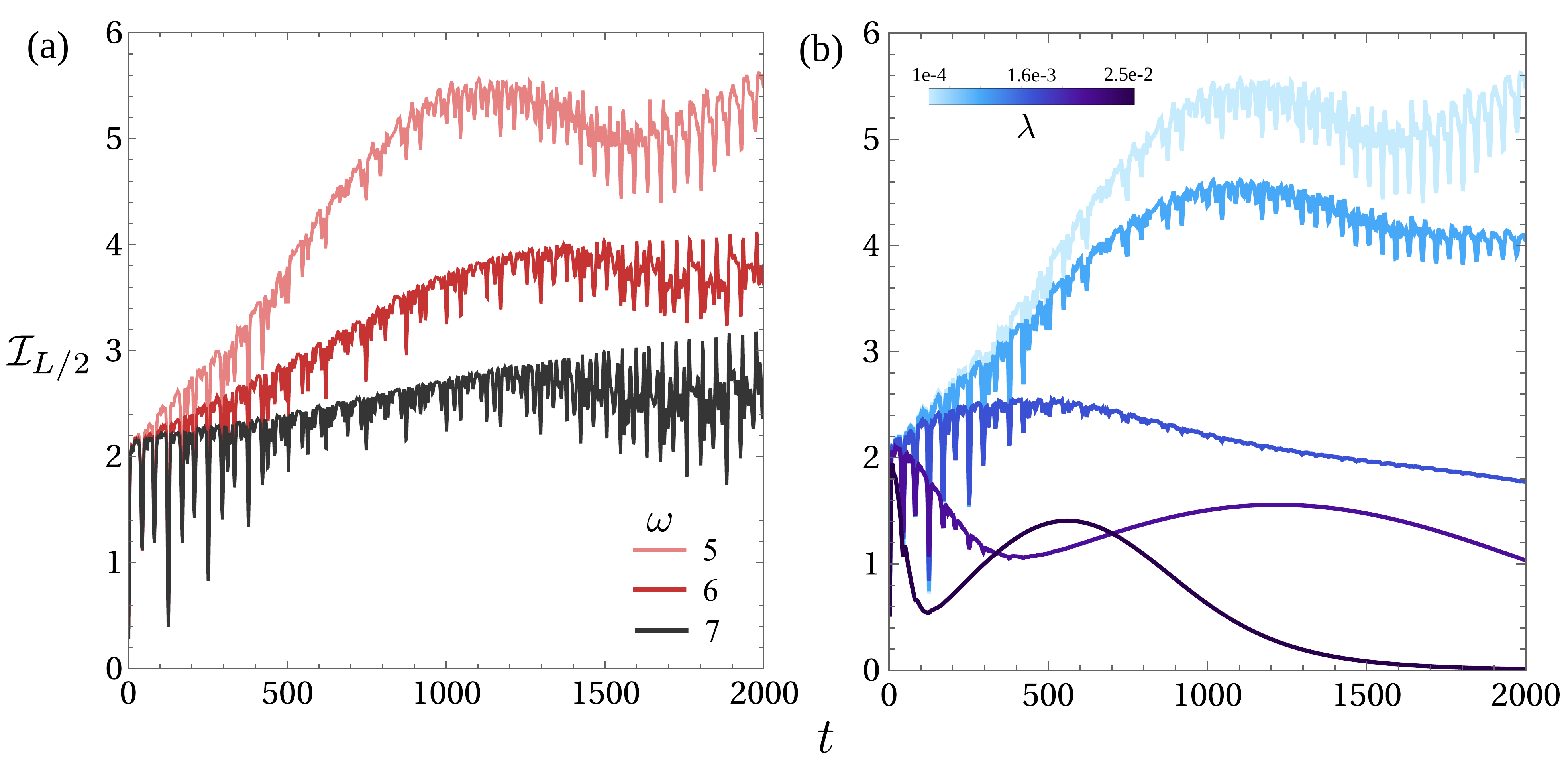}
	\caption{(a) Bipartite mutual information for closed spin chain at different drive frequencies. (b) Mutual information at $\omega=5$ with different dephasing strengths. \label{fig4}}
\end{figure}

    Beside the magnetization, we also compute the bipartite mutual information to understand which mechanism is responsible for the DTC decay. A subsystem of the spin chain either the rest of the chain (internal coupling) or the thermal bath (external coupling). By defining the mutual information between the two halves of chain $\mathcal{I}_{L/2} = S_A +S_B-S_{AB}$ with $A$ and $B$ being the two halves and $S_A$ being the Renyi entropy of the subsystem $A$. In Fig.~\ref{fig4}(a), we demonstrate the prethermal physics in closed systems. Under the thermalization generated by internal coupling, the mutual information always increases until saturation with slower rate for larger $\omega$. On the other hand, if bath coupling dominates, the system becomes purely classical with vanished mutual information. In Fig.~\ref{fig4}(b), we compare the behavior of $\mathcal{I}$ within $\tau_{DTC}$. For low $\lambda$, thermalization is primarily driven by the internal coupling, characterized by the increasing of $\mathcal{I}$ with time. However, the rate decreases, consistent with the prolonged $\tau_{\text{DTC}}$. When the dephasing is sufficiently strong, the bath coupling takes over the thermalization, as shown in the decreasing mutual information. For $\omega=5$, the transition happens around $\lambda\approx 0.002$ which also coincides with the peak in Fig.~\ref{fig2}(c). As such, the nonmonotonic behavior reported in our paper can be associated with the transition from the quantum to classical dynamics.
    
\begin{figure}
	\centering
	\includegraphics[width=0.5\textwidth]{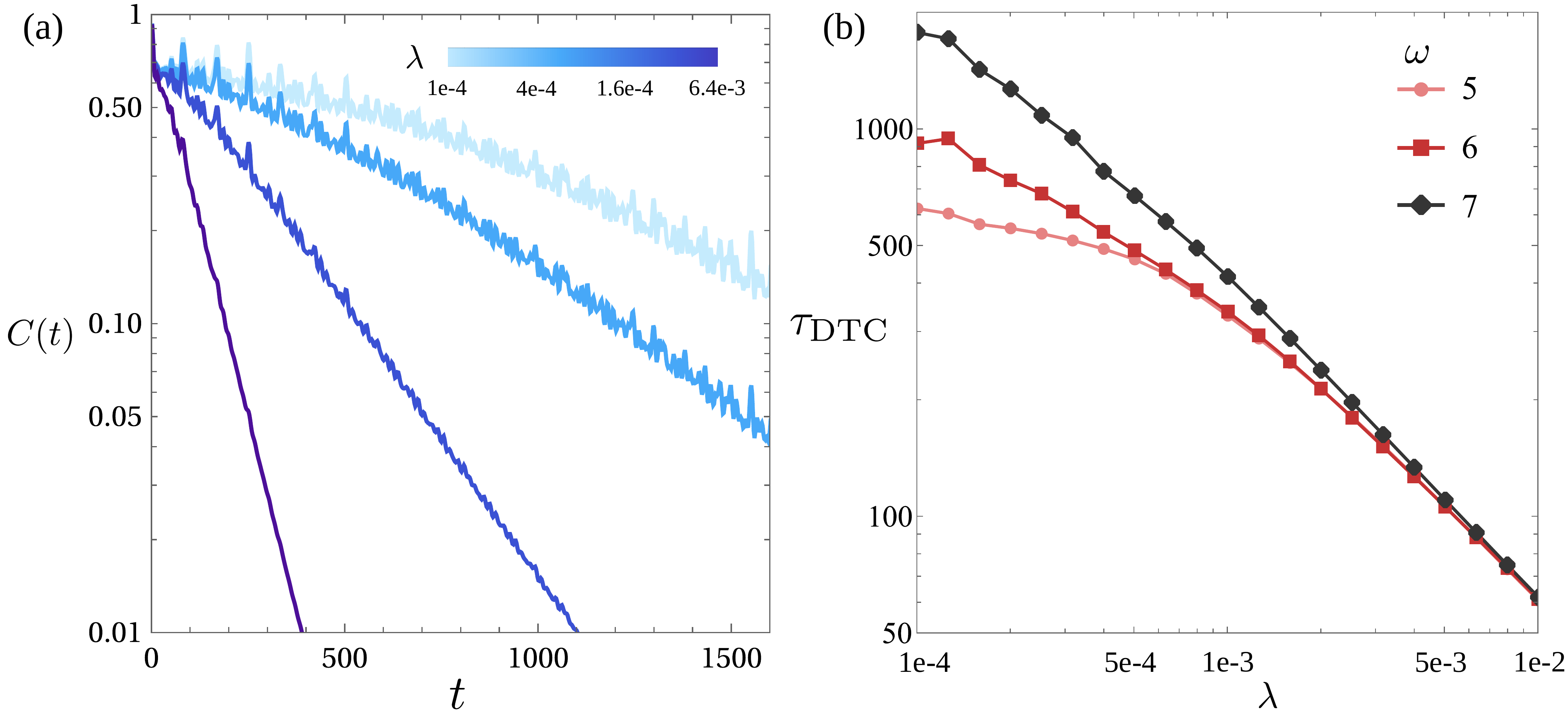}
	\caption{DTC signal with dephasing strength (a) and DTC lifetime with applied frequency (b) under $s_x-$dephasing noise. In this case, the gain regime is almost invisible. \label{fig3}}
\end{figure}

    \textit{$S_x-$dephasing noise - } Compared to the previous case of noise along the $z-$direction, the decreasing trend is much more visible while the increasing one is negligible [see Fig.\ref{fig3}]. In Fig.~\ref{fig3}(b), we show the scaling with respect to frequency. The invariance against the applied frequency proves that $\tau_{\text{DTC}} = \tau_{\text{SSB}}$, consistent with our general result. The reason being the decay of SSB under $S_x-$noise is much faster than under $S_z$ noise, making the observed variation of $\tau_{\text{DTC}}$ with respect to $\lambda$ being heavily biased toward the decreasing phase.

    \textit{Conclusion - } We establish the general trends of the dissipative prethermal DTC in the presence of an environmental coupling. The physics can be divided into two phases. The first increasing phase where the DTC lifetime increases with the environmental coupling, accompanied by an exponential dependence on the frequency. The second decreasing phase, by contrast, has the DTC lifetime shortened with increasing noise strength and has no frequency dependence. We note that in most of the DTC literature in the presence of bath, the instability from bath coupling is emphasized \cite{Lazarides2017,Lazarides2020}. The stabilizing effect reported in our work is surprising, but can be hard to observe if the form of disspipation is chosen incorrectly. One situation where the environmental coupling is beneficial is the classical DTC where the bath manifests as a damping force and noise \cite{Yao2020}. In our case, the mechanism underlying the stable branch is also the damping of nonsymmetric error accumulation manifesting in the decreasing Lieb-Robinson velocity. It is therefore interesting to draw some connection between the classical and quantum DTC.

    The physics reported in our paper can be realized in available quantum devices using stochastic trajectories. In particular, each trajectory is subjected to a randomized Ising field as
	\begin{equation}
	  H_{ST}(t) = H(t) + \sum_i \epsilon_i(t) C^i, 
	\end{equation}
    where $C$ is either $S_z$ or $S_x$ and $\epsilon$ is a random number chosen from a Gaussian distribution so that $\overline{\epsilon_i(t)}=0$ and $\overline{\epsilon_i(t)\epsilon_j(t')} = \lambda \Delta t \delta_{i,j}\delta_{t,t'}$. Even though each trajectory is entirely Hermitian, but the effect of the dephasing noise shows up in the ensemble over random configurations.
    
     \begin{acknowledgments}
	\textit{Acknowledgments - }
	This work is supported by and Laboratory for Physical Sciences. This work is also supported by the High Performance Computing Center (HPCC) at the University of Maryland. The authors are grateful to Minh Tran for helpful discussions.
\end{acknowledgments}

\bibliographystyle{apsrev4-2}
\bibliography{Time crystal}
\end{document}


\title{Supplemental Materials for ``Dissipative prethermal discrete time crystal''}
\maketitle
\section{Prethermalization in dissipative systems}
\subsection{Supervector and superoperator formalism}
In this section, we translate the formalism from the density matrix/ quantum channel language to the supervector/ superoperator
\begin{equation}
\begin{split}
&\rho \to \kket{\rho} \\
&\rho'=D[\rho]=A\rho B \to \kket{\rho'} = \hat{D}\kket{\rho} \text{ with } \hat{D}=A\otimes B^T\\
& \text{Tr}(\rho.\nu) \to \bbra{\rho} \nu \rrangle
\end{split}
\end{equation}
Within this formalism, we can obtain several identities
\begin{equation}
\begin{split}
& \hat{D}^\dagger\kket{\mathbbm{1}} = 0 \leftrightarrow \bbra{\mathbbm{1}}e^{\hat{D}}\kket{\rho} = \bbra{\mathbbm{1}} \rho \rrangle \leftrightarrow e^{D} \text{ is a trace-preserving map}\\
& \text{$X$ is a completely positive trace-preserving map } \to X[\mathbbm{1}]=\sum_i V_i\mathbbm{1}V_i^\dagger =\mathbbm{1} \leftrightarrow \hat{X}\kket{\mathbbm{1}} = \kket{\mathbbm{1}} 
\end{split}
\end{equation}
We mention that if $e^{D_1}$ and $e^{D_2}$ are CPT, then $e^{D_1+D_2}$ is also CPT (because of the positivity this is not true generally for $e^{D_1-D_2}$). We also introduce the superoperator norm to quantify that operator's action.
\begin{equation}
\norm{\hat{A}} = \sup_{\kket{\rho}} \frac{\left|\hat{A}\kket{\rho}\right|}{\left|\kket{\rho} \right|} = \sup_\rho \frac{\sqrt{\text{Tr}\left(A[\rho]^\dagger A[\rho]\right)}}{\sqrt{\text{Tr}\left(\rho^\dagger \rho\right)}}.
\end{equation}
For $\hat A = \sum_{Z} \hat{A}_Z$ where $Z$ is the superoperator support 
\begin{equation}\label{norm1}
\norm{\hat{A}}_{\kappa_n} = \sup_j \sum_{Z \ni j}e^{\kappa_n \left|Z \right|}\norm{\hat{A}_Z} \text{ with }\kappa_n \equiv \frac{\kappa_1}{1+\log n}.
\end{equation}
Here, $\kappa>0$ counteracts the exponential suppression of large-support operators.
For long-range Linbladian, the support $Z$ of an operator string $\hat{A}_{Z,R}$ contains extra information of disconnected clusters separated by at most $R$. To keep track of this separation, we can use the two-parameter norm as in Ref.~\cite{Machado2020}
\begin{equation}\label{norm2}
\norm{\hat{A}}_{\eta,\kappa} = \sup_j \sum_R R^{\eta} \sum_{Z\ni j} e^{\kappa |Z|} \norm{\hat{A}_{Z,R}}.
\end{equation}
Similar to the one-parameter norm, $\eta>0$ characterizes the algebraic decay of the long-range interaction. If the operator is time-dependent, the norm is defined as the average of the instantaneous norm over a cycle.

\subsection{Iterative optimization}
We first review the prethermal physics in closed systems with short and long-range interactions. In both cases, the Floquet operator generated by a temporally periodic Hamiltonian $H(t+T)=H(t)$
\begin{equation}
U_f = \mathcal{T}\text{exp}\left( -i\int_0^T H(t)dt \right),
\end{equation}
under some local basis rotation, can be approximated by an evolution under a time-independent Hamiltonian. The accuracy of this approximation improves exponentially with the drive frequency.\\

\textit{\textbf{Theorem 1:}} Assume $H(t)$ is short-ranged and $H(t)=H_0(t)+V(t)$ where $H_0$ contains only onsite terms, $X\equiv \mathcal{T}\exp\left(-i\int_0^T H_0(t)\right)$ satisfies $X^N=1$ with some integer $N$, and $\gamma=\norm{V}_1$ satisfies 
\begin{equation}
\gamma T \le \frac{\mu \kappa_1^2}{N+3} \quad \mu\approx 0.14.
\end{equation}
Under a unitary transform $\mathcal{U}=e^{-iA_1}\dots e^{-iA_n}$ with quasi-local $A_n$,
\begin{equation}
\mathcal{U}U_f\mathcal{U}^\dagger = X\mathcal{T}\exp\left[-i\int_0^T (D^*+E^*+V^*(t))dt \right],
\end{equation} 
where $\left[D,X^*\right]=0$; $D$ and $E$ are time-independent while $V^*(t)$ integrates to zero over one period; and
\begin{equation}
\norm{E}_{\kappa_{n*}},\norm{V}_{\kappa_{n*}} \le \frac{\gamma}{2^{n*}}~\text{ with } n^* = \frac{\gamma_0/\gamma}{\left[1+\log(\gamma_0/\gamma)\right]^3},~\gamma_0 = \frac{\kappa_1^2}{72(N+3)(N+4)T}. 
\end{equation}
Furthermore, $\mathcal{U}$ approaches the identity operator in the large-$\omega$ limit in the sense that
\begin{equation}
\norm{\mathcal{U}\Phi\mathcal{U}^\dagger}_{\kappa_{n^*}} \le C\gamma T \norm{\Phi}_{\kappa_1}
\end{equation} 
for any operator $\Phi$.

Proof: See Ref.~\cite{Else2017}\\

\textit{\textbf{Theorem 2:}} Assume $H(t)$ has terms with power-law decay, i.e., $\norm{H_{i,j}} \le C/\text{dist}(i,j)^\alpha$ and 
\begin{equation}
U_f = X\mathcal{T}\exp\left[-i\int_0^T (D+E+V(t))dt \right].
\end{equation}
with $[D,X]=0$, $D$ and $E$ are independent, and $V(t)$ integrates to zero.  With fixed $\kappa, \eta$, and $0<\mathcal{E}<1$ and $\gamma = \max\{\norm{E}_{\kappa,\eta}, \norm{E}_{\kappa,\eta}, \norm{V}_{\kappa,\eta}\}$, there exists a unitary transform $\mathcal{U}$ and constants $C_1,\dots,C_5>0$ such that if $\gamma T \le C_1$ then
\begin{equation}
\mathcal{U}U_f\mathcal{U}^\dagger = X\mathcal{T}\exp\left[-i\int_0^T (D^*+E^*+V^*(t))dt \right],
\end{equation}
where
\begin{equation}
\norm{D-D^*}_{\kappa^*,\eta^*} \le C_3 \gamma^2T;~\norm{E^*}_{\kappa^*,\eta^*},\norm{V^*}_{\kappa^*,\eta^*} \le \frac{C_2 \gamma^2 T}{2^{n^*}} 
\end{equation}
and
\begin{equation}
\kappa^* = \mathcal{E}\kappa,~\eta^*=\mathcal{E}\eta,~n^* = \frac{C_4}{\gamma T}.
\end{equation}
Moreover, the unitary transformation $\mathcal{U}$ is close to the identity transformation in the precise sense that
\begin{equation}
\norm{\mathcal{U}\Phi\mathcal{U}^\dagger - \Phi}_{\kappa^*,\eta^*} \le C_5\gamma T\norm{\Phi}_{\kappa,\eta}.
\end{equation}
for any range-indexed operator $\Phi$.

Proof: See Ref.~\cite{Machado2020}\\

\textit{\textbf{Theorem 3:}} Consider a Linbladian $\mathcal{L}(t)$ so that $\partial_t\rho = \mathcal{L}(t)\rho$ and $\mathcal{L}(t)=\mathcal{L}(t+L)$. By promoting to the double Hilbert space, i.e. $\mathcal{L}\to\hat{\mathcal{L}}$, we define the Floquet superoperator similar to the Hermitian case
\begin{equation}
\hat{U}_f=\hat{X} \mathcal{T} e^{\int_0^T \hat{\cl}(s)ds}.
\end{equation} 
We also assume the Hermitian part of the Linbladian contains $H_0(t)$ similar to Theorem 1 and 2 that gives rise to the emergent symmetry operator $X$ in the original Hilbert space. The action of the symmetry on a density matrix is given by the mapping $\rho \to X\rho X^\dagger$, corresponding to a superoperator $\hat{X}$ in the double Hilbert space.
There exist a trace-preserving map $e^{\hat{A}^*}$ such that
\begin{equation}
e^{\hat{A}^*} \hat{U}_f e^{-\hat{A}^*} = \hat{X}\mathcal{T}\exp\left[\int_0^T (\hat{D}^*+\hat{E}^*+\hat{V}^*(s))ds\right],
\end{equation}
where $[\hat{D}^*,\hat{X}]=0$; $\hat{D}^*$ and $\hat{E}^*$ are time-independent; $\hat{V}^*(t)$ integrates to zero. Furthermore, $\hat{D}^*$ is a CPT generator, while $\hat{A},\hat{E}^*,\hat{D}^*$ are trace-preserving map generators. The norm of the error terms $\hat{E}^*,\hat{D}^*$ are suppressed exponentially by the drive frequency by Theorem 1 or 2 depending on whether the interaction is short or long-ranged.\\

Proof: We prove Theorem 3 by the iterative method, following closely Refs.~\cite{Abanin2017,Else2017,Machado2020}. By mapping the Lindbladian dynamics to the evolution of the superoperator, we can prove the prethermalization in open systems exactly in the same manner as in closed systems. Therefore, we only demonstrate the short-range Lindbladian case here, focusing on proving that the symmetric superoperator $\hat{D}^*$ after iterative optimization generates a legitimate density matrix transformation, i.e., a complete positive trace-preserving map (CPT). We first assume that at some step $n$, the evolution over a driving period can be written in this form
\begin{equation}
\hat{U}(0,T)=\hat{X} \tau e^{\int_0^T \cl(s)ds}
\end{equation}
We define $\hat{F}=T^{-1}\int_0^T\cl(s)ds$, $\hat{F}$ is a CPT generator because $\cl(s)$ is one at any $0\le s \le T$. We can then symmetrize this superoperator
\begin{equation}
\hat{D}_n = \frac{1}{N}\sum_{k=0}^{N-1} \hat{X}^k \hat{F} \hat{X}^{-k}
\end{equation}
This symmetrized $\hat{D}_n$ is also a CPT generator. The residue $\hat{E}_n$ if we assume to a CPT, then $-\hat{E}_n = N^{-1}\sum_{k=1}^{n_1}\hat{X}^k\hat{E}_n\hat{X}^{-k}$ is also a CPT generator. That means $\hat{E}_n$ only have purely imaginary eigenvalues. The time dependent superoperator is denoted at  $\hat{V}$ satisfying $\int_0^T \hat{V}(s)ds = 0$. We now apply the transformation
\begin{equation}
\hat{U}' = e^{-\hat{A}} \hat{U} e^{\hat{A}} =  \hat{X} \tau (\hat{X}^\dagger e^{-\hat{A}}\hat{X})e^{\int_0^T \cl(s)ds} e^{\hat{A}} = \hat{X}\tau e^{\int_0^T \cl'(s)ds}
\end{equation}
with
\begin{equation}
\cl'(s)=\begin{cases}
a^{-1}(\hat{A}) & s<=a\\
(1-2a)^{-1}\cl\left( \frac{s-a}{1-2a} \right) & a < s <= 1-a \\
a^{-1}\left(-\hat{X}^\dagger \hat{A} \hat{X} \right) & s>1-a
\end{cases}.
\end{equation}
We want to choose the map $\hat{A}$ such that $\hat{D}'_n=\hat{D}_n$ and $\hat{E}'_n=0$. It easy to see that with the choice
\begin{equation}
\hat{A}_n = \frac{1}{N}\sum_{k=0}^{N-1}\sum_{p=0}^k \hat{X}^{-p}\hat{E}_n\hat{X}^p,
\end{equation}
then $\hat{E}_n - \hat{X}^\dagger \hat{A}_n\hat{X}+\hat{A}_n =0 $. We also note that if $\text{Tr}\left(E_n[\rho]\right)=0$ for all density matrices $\rho$ then $\text{Tr}\left(A[\rho]\right)=0$ and $e^{\hat{A}}$ is a trace-preserving map. The time-dependent superoperator is given by
\begin{equation}
\hat{V}'_n(s)=\begin{cases}
a^{-1}(\hat{A}_n) - \hat{D}_n & s<=a\\
(1-2a)^{-1}\left[2a\hat{D}_n+\hat{E}_n+\hat{V_n}\left(\frac{s-a}{1-2a}\right)\right] & a < s <= 1-a \\
a^{-1}\left(-\hat{X}^\dagger \hat{A}_n \hat{X} \right) -\hat{D}_n & s>1-a
\end{cases}.
\end{equation}
By this procedure, even though $\hat{E}_n$ is eliminated, $\hat{V}'_n$ is large by the virtue of $\hat{D}_n$. We now find an alternative $\cl_{n+1}$  that produces the same evolution over one period $T$ as $\cl'_n$ so that $\hat{V}_{n+1}$ is reduced. We assume the form
\begin{equation}
\tau e^{\int_0^t \cl_{n+1}(s)ds} = e^{-\hat{K}_n(t)}\tau e^{\int_0^t \cl'_n(s')ds'}
\end{equation}
Our condition is satisfied if $\hat{K}_n(0)=\hat{K}_n(T)=0$. Within this construction
\begin{equation}
\cl_{n+1}(t) = e^{-\hat{K}_n(t)}\cl'_n(t)e^{\hat{K}_n(t)} - e^{-\hat{K}_n(t)}\partial_t e^{\hat{K}_n(t)}
\end{equation}
If we choose $\hat{K}$ such that $\partial_t\hat{K}_n(t) = \hat{V}_n(t)$, then $\hat{K}_n(0)=\hat{K}_n(T)=0$ because $\int_0^T \hat{V}(s)ds=0$. Then
\begin{equation}
\begin{split}
\hat{D}_{n+1}+\hat{E}_{n+1}-\hat{D}_n &= \frac{1}{T}\int_0^T \left[e^{-\hat{K}_n(s)} \hat{D}_n e^{\hat{K}_n(s)} - \hat{D}_n\right]ds + \frac{1}{T}\int_0^T \left[e^{-\hat{K}_n(s)} \hat{V}_n(s) e^{\hat{K}_n(s)} - \hat{V}_n(s)\right]ds \\
& - \frac{1}{T}\int_0^T\int_0^1\left[e^{-s'\hat{K}_n(s)} \hat{V}_n(s) e^{s'\hat{K}_n(s)} - \hat{V}_n(s)\right]ds'ds
\end{split}
\end{equation}
Here, we make use of a bound \cite{Abanin2017}
\begin{equation}
\norm{e^{-K}Ae^{K} - A}_{n+1} \le \frac{18}{\delta \kappa_n\kappa_{n+1}} \norm{A}_n\norm{K}_n
\end{equation}
that is valid for both Hermitian and non-Hermitian matrices as long as $3\norm{K}_n\le \delta \kappa_n\equiv \kappa_n-\kappa_{n+1}$. For conciseness, we denote $\norm{.}_{\kappa_n}$ as $\norm{.}_n$. We remind that $\norm{\sum_j A_j}_n \le \sum_j\norm{A_j}_n$. As a result,
\begin{equation}
\norm{\hat{D}_{n+1}+\hat{E}_{n+1}-\hat{D}_n}_{n+1} \le \frac{18}{\delta \kappa_n\kappa_{n+1}}\norm{K}_n\left( \norm{D_n}_n + \norm{V_n}/2 \right)
\end{equation} 
Also
\begin{equation}
\hat{V}_{n+1}(t) = e^{-\hat{K}_n(t)} \hat{D}_n e^{\hat{K}_n(t)}-(\hat{D}_{n+1}+\hat{E}_{n+1}) + e^{-\hat{K}_n(t)}\hat{V}_n(t)e^{\hat{K}_n(t)} - \int_0^1 ds e^{-s\hat{K}_n(t)}\hat{V}_n(t)e^{s\hat{K}_n(t)}
\end{equation}
leading to the norm
\begin{equation}
\norm{\hat{V}_{n+1}}_{n+1} \le \frac{18}{\delta \kappa_n\kappa_{n+1}}\norm{K}_n\left( \norm{D_n}_n + \norm{V_n} \right)
\end{equation}
It is straightforward to see that $\hat{\cl}_{n+1}$ is a trace-preserving map generator, but we do not have a rigorous way to show it is a CPT generator. However, by sending $a\to 0$, to the first order of $\hat{K}_n$
\begin{equation}
\hat{\cl}_{n+1}(s)= \hat{D}_n + [\hat{D}_n,\hat{K}_n(s)] - \hat{V}_n(s).
\end{equation}
Because $\hat{D}_n$ is a CPT generator and the other parts are small, $\hat{\cl}_{n+1}(s)$ is most likely also a CPT generator. Iterating the procedure, for any $n$, $e^{\hat{A}_n}$ is always a trace preserving map and $\hat{\cl}_n(s),\hat{D}_n$ are always CPT generator. By applying the same induction method, we obtain a similar bound as in the unitary case.

\section{Effect of dissipation on the prethermal period}
\subsection{Slow heating and approximation error}
We show the error by approximating the full dynamic by only the time-independent symmetric Liouvillian $\hat{D}$
\begin{equation}
\delta(t) = \bbra{\mathbbm{1}} \hat{O} e^{\hat{A}}\hat{U}(0,t)e^{-\hat{A}} - \hat{O} e^{\hat{D}t} \kket{\psi}
\end{equation}
We use the Duhamel's formula if $\partial_t U(t) = [A+B(t)]U(t)$
\begin{equation}
U(t) = e^{At} + \int_0^t e^{As}B(t-s)U(t-s)ds
\end{equation}
As a result,
\begin{equation}
\delta(t) = \int_{0}^{t} \bbra{\mathbbm{1}}  O e^{s\hat{D}}\hat{V}(t-s) \kket{\rho(t-s)} ds
\end{equation}
By assuming $\hat{V}=V\otimes\mathbbm{1} -\mathbbm{1}\otimes V^T$, i.e., $\hat{V}$ only contains the coherent part, we can translate the error bound to a more familiar form (we drop the argument $t-s$ from $\hat{V}$ and $\rho$ for conciseness)
\begin{equation}\label{errorbound}
\delta(t) = \int_0^t \text{Tr}\left(Oe^{Ds}[V\rho-\rho V]\right) ds = \int_0^t \text{Tr}\left(\left[V,e^{D^\dagger s}[O]\right]\rho\right) ds
\end{equation}
which is reminiscent of the Lieb-Robison bound in a dissipative system. 	From Eq.~\eqref{errorbound}, we can derive the prethermal period on several occasions. Generally, the form of Lieb-Robison bound does not change between closed and open systems and only depends on the range of interaction. For a Liouvillian $\hat{V}(Y)$ with support on $Y$ and an operator $O(X)$ with support on $X$ such that $X\cap Y =\emptyset$ and $d=d(X,Y)$ is the distance between the two supports defined in some metric, the Lieb-Robinson is given by
\begin{equation}
\norm{\hat{V}(Y) e^{D^\dagger t}[O(X)]} \le C\norm{\hat{V}}\norm{O}\left| X\right| \left| Y \right|  e^{\mu(vt-d)}
\end{equation}
for short-range Lindbladian and 
\begin{equation}
\norm{\hat{V}(Y) e^{D^\dagger t}[O(X)]} \le C\norm{\hat{V}}\norm{O}\left| X\right| \left| Y \right| \frac{e^{vt} - 1}{(1+d)^\alpha}
\end{equation}
for long-range Lindbladian with interaction $\sim 1/r^{\alpha}$ \cite{Sweke2019,Guo2021}. Here, $C$, $\mu$, and $v$ are constant, with the third one called Lieb-Robision velocity. The noise we use in the main text is strictly onsite, so at first sight, it should not affect the propagating velocity $v$. However, Ref.~\cite{Descamps2013} suggests a scenario where propagation speed is damped exponentially with time under the bath dissipation. We expect that this effect slows the growth of $\delta(t)$ in Eq.~\eqref{errorbound}, effectively prolonging the prethermal period. However, in Ref.~\cite{Descamps2013}, the role of the bath coupling is not shown explicitly. By the perturbative method, we show that the damping rate of the Lieb-Robison bound in the dephasing noise case is proportional to the noise strength.

\subsection{Noise-dampened Lieb-Robison velocity}

\textit{\textbf{Theorem 4:}} Given a local Lindbladian $\mathcal{L} = \sum_X I_X$ with the conjugate $I_X^*[1]=0$ and $P_X$ as the projector on $\text{Ker}(I_X+I_X^*)$, if
\begin{equation}\label{condition}
P_Y I_X (1-P_Y) =0,~P_YP_X(1-P_Y) = 0 \quad \forall X,Y
\end{equation}
then $\forall A_{Z_A}$ and $B_{Z_B}$ so that $Z_A\cap Z_B = \emptyset$
\begin{equation}
\norm{[e^{tL}[A],B]} \le C\norm{A}\norm{B}\min\left(|Z_A|,|Z_B|\right)\left(\exp \left[\int_0^t v(s)ds\right] -1\right)\exp[-\mu d(A,B)]
\end{equation}
with decelerated velocity
\begin{equation}
v(s)= C\left(\alpha+\beta e^{-\delta s}\right)
\end{equation}
where
\begin{equation}\label{eq41}
\delta = - \sup_X E_{I_X+I_X^*/2},\quad\alpha=\sup_X \norm{I_XP_X}/\norm{I_X},\quad \beta=\sup_X \norm{I_X(1-P_X)}/\norm{I_X}
\end{equation}
and $E_T$ is the largest non-zero eigenvalue of $T$.

Proof: See Ref.~\cite{Descamps2013}

The intuition behind Theorem 4 is that as the support grows larger, the suppression rate caused by dissipation is stronger, effectively slowing down the information propagation. 
In our model, given a Pauli string, after a time step $\Delta t$, the dephasing, for example, along the $z$-basis sends $O\to e^{-\lambda(N_x+N_y)/2}O$ with $N_{x,y,z}$ being the number of $\sigma_{x,y,z}$ in the string and $\lambda$ is the dephasing strength used in the main text. If we assume the $N_x\approx N_y\approx N_z$, then $e^{-\lambda|Z_O|/3}$ with $Z_O$ being the support of the Pauli string $O$. Equivalently, we can replace the dephasing noise along a preferred direction with the isotropic dissipation
\begin{equation}
\mathcal{L}_D[.] =  \frac{\lambda}{3}\sum_i\left(s_x^i[.]s_x^i+s_y^i[.]s_y^i+s_z^i[.]s_z^i -3[.] \right)
\end{equation}
It is obvious to see that in this case $\text{Ker}(I_X+I_X^*)=1_X$ and conditions~\eqref{condition} are satisfied because $I_X[1_X]=0$ (the quantum jump operators are all Hermitian). Equation~\eqref{eq41} yields $\delta =\lambda/3$, $\alpha=0$ and $\beta\sim \mathcal{O}(1)$. Using Theorem 4, the Lieb-Robison velocity becomes $v(s) = v_0 e^{-\lambda t/3}$.

We first apply the damped Lieb-Robinson velocity to Eq.~\eqref{errorbound} in the short-range Lindbladian. In the presence of bath coupling, the size of the support is modified to $\left|O \right| = s_0 + 3v\left(1-e^{-\lambda t/3}\right)/\lambda$ with $s_0$ being the initial size of the support at $t=0$. Consequently, Eq.~\eqref{errorbound} yields
\begin{equation}
\delta(t) \approx \eta 2^{-n^*}\int_0^t \left[ s_0 + \frac{3}{\lambda}\left(1-e^{-\lambda s/3}\right) \right] ds =  \eta 2^{-n^*}\frac{3v\left(3e^{-\lambda t/3} +\lambda t -3\right)}{\lambda^2}
\end{equation}
In the dissipationless limit $\lambda\to 0$, it is easy to see that $\tau_{\text{prethermal}}^0 \propto 2^{n^*/2} \sim e^{\omega/(2\norm{\hat{V}}_1)}$. From our numerical simulation, $\tau_{\text{prethermal}}^0\sim \mathcal{O}(10^2)-\mathcal{O}(10^3)$ so $\lambda\tau_{\text{prethermal}}$ is not a negligible number even though $\lambda\ll J \ll \omega$. To the first order of $\lambda$, the prethermal period in the presence of a bath can be estimated by
\begin{equation}\label{SR}
\tau_{\text{prethermal}}^\lambda =\tau_{\text{prethermal}}^0 \left(1- \frac{\lambda \tau_{\text{prethermal}}^0}{9}\right)^{-1},
\end{equation}
explaining the visible effect of the bath coupling even though its magnitude is much smaller than any other energy scales. 

In the case of a closed long-range interacting system, the accumulated error is more complex due to the two-parameter norm and given by
\begin{equation}
\delta(t)\propto 2^{-n^*}t\left[K_3(1+\tau^{d/(1-\xi)}) + K_4 (\tau+\tau^\beta) \right]
\end{equation}
with $\tau=vt$ and other constants defined in Ref.~\cite{Machado2020}. It is difficult to obtain explicitly the prethermal range from this equation. However, estimating the relation similar to Eq.~\eqref{SR} is easier. In general, including the long-range case, the prethermalization lifetime in a closed system can be obtained by solving
\begin{equation}
\int_0^{\tau_{\text{prethermal}}^0} f(vs)ds  \sim 2^{n^*}.
\end{equation}
Here $f$ is the corresponding Lieb-Robinson bound, and $v$ is the Lieb-Robinson velocity. A rescaling $v\to va$ leads to $\tau_{\text{prethermal}}\to a^{-1}\tau_{\text{prethermal}}\left(2^{n^*}a^{-1}\right)$. The decelerating effect generated by the environment modifies $vs \to 3v(1-e^{\lambda s/3})/\lambda \approx vs(1-\lambda s/6)$. As a result, we can generalize Eq.~\eqref{SR} as $\tau_{\text{prethermal}}^\lambda =\tau_{\text{prethermal}}^0 \left(1-C \lambda \tau_{\text{prethermal}}^0\right)^{-1}$
with $C$ being an $\mathcal{O}(1)$ constant.

\section{Effect of dissipation on the quasi-spontaneous symmetry breaking}

\begin{figure}
	\centering
	\includegraphics[width=0.55\textwidth]{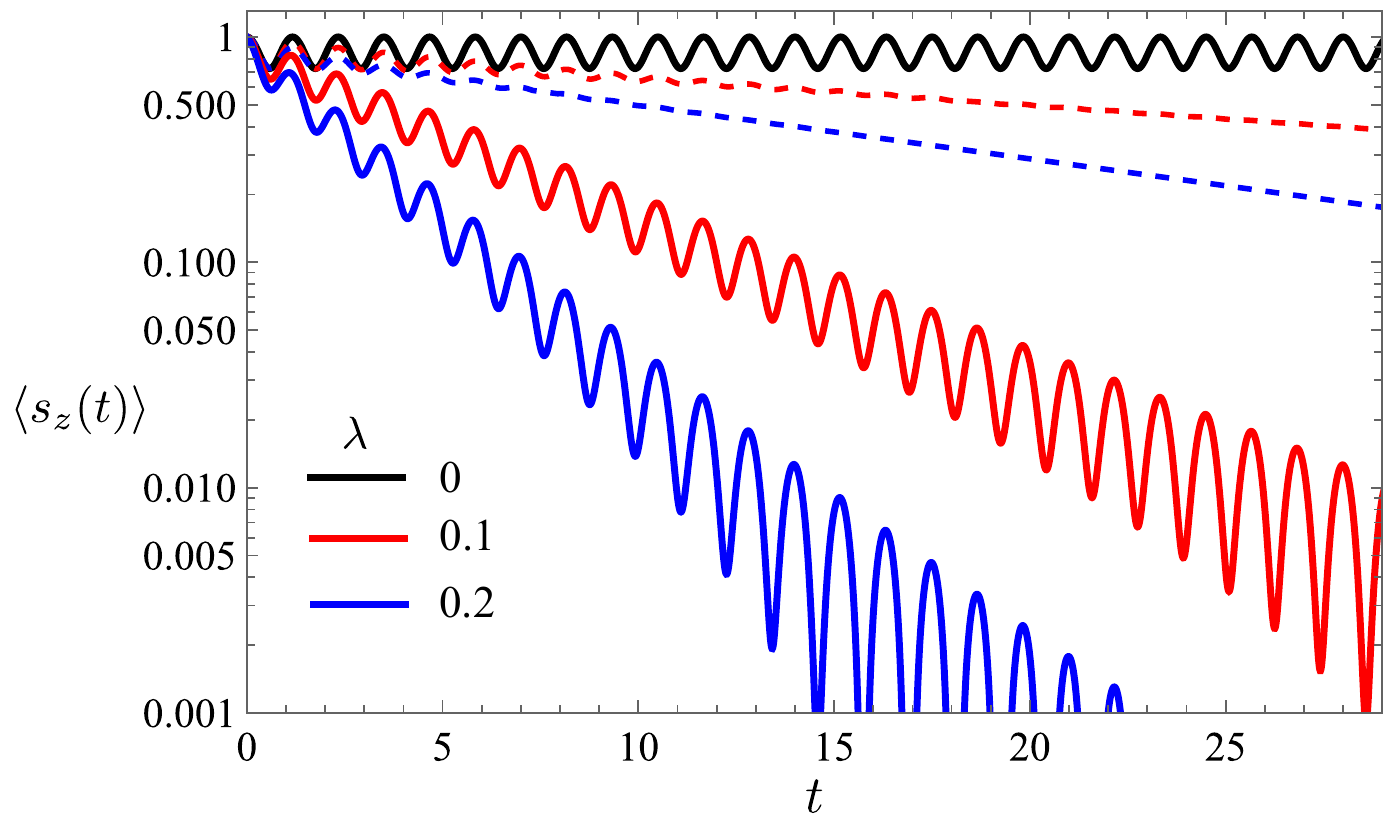}
	\caption{Evolution of $\mean{s_z}$ obtained from numerical simulation with $J=1$, $U=5$. The solid (dashed) lines represent the system coupled to $s_x$ ($s_z$)-dephasing noise. \label{figSM1}}
\end{figure}

In a closed system, a spontaneous symmetry breaking is protected by energy conservation, i.e., a spin-flip excitation costs energy and thus is short-lived, while simultaneous spin-flipping throughout the system is exponentially suppressed. For an open system, energy can be exchanged with the bath to facilitate a single excitation. In addition, the spontaneous symmetry breaking is stabilized by the long-range interaction in 1D, while the dissipation we use is strictly onsite and thus tends to disrupt long-range correlation. Therefore, we expect the lifetime of a quasi-SSB state to decrease with stronger bath coupling. In fact, in a closed system, the ground state energy splitting $\sim e^{-\beta L}$ with $L$ being the system size, leading to the $\tau_{\text{SSB}}^\text{closed}\sim e^{\beta L}$. On the other hand, in a closed system with local dissipation, the lifetime of the quasi-SSB state is bounded from below by $\sim L^{d/(d+1)}=L^{1/2}$ in 1D \cite{Wilming2005}. As such, the finite-size effect is much more severe in the open system and should be taken into account.

The power-law scaling of $\tau_{\text{SSB}}$ derived in Ref.~\cite{Wilming2005} is also based on the Lieb-Robinson bound of perturbation propagation. However, the dependence on the strength and form of the dissipation is not transparent in their results. Our numerical results in the main text show that the dephasing noises along the $z-$ and $x-$ directions show deviating effect. For this reason, we analytically study a model of a single spin in the presence of a Zeeman field (representing a correlation effect with neighboring spins) 
\begin{equation}
\cl [\rho] = -i \left[\begin{pmatrix}
0 & J\\ J & U
\end{pmatrix}, \rho \right] + \lambda\left(\sigma_{x,z}\rho\sigma_{x,z}-\rho\right).
\end{equation}
Here, $\rho$ is in the $\{\ket{\uparrow},\ket{\downarrow}\}$, $J$ is spin-flipping magnitude, $U$ is energy cost of the spin-flip excitation and $\sigma_{x,z}$ are Pauli matrices. We assume $J\ll \lambda \ll U$ so that the dissipation is much slower than the coherent evolution of the spin. We initialize the density matrix at $\rho(0) = \ket{\uparrow}\bra{\uparrow}$ and measure the evolution of the spin along the $z-$direction $\mean{s_z(t)}=\text{Tr}\left( s_z e^{\cl t}[\rho] \right)$. In the limit $\lambda\to 0$, the system sustains undamped oscillations so that $1\ge \mean{s_z(t)} \ge 1-8J^2/(4J^2+U^2)$.

We now study the role of dissipation perturbatively. It is more convenient to write the Lindbladian in superoperator form $\hat{\cl} = \hat{\cl}_0 + \hat{\cl}_{Dz, Dx}$ and perform time-dependent perturbation theory
\begin{equation}
\hat{\cl}_0= \begin{pmatrix}
0 & iJ & -iJ & 0\\
iJ & iU & 0 & -iJ\\
-iJ & 0 & -iU & iJ \\
0 & -iJ & iJ & 0 
\end{pmatrix}, \quad \hat{\cl}_{Dx} = \lambda 
\begin{pmatrix}
-1 & 0 & 0 & 1\\
0 & -1 & 1 & 0\\
0 &  1 & -1 & 0\\
1 & 0 & 0 & -1
\end{pmatrix}, \quad \hat{\cl}_{Dz} = \lambda 
\begin{pmatrix}
0 & 0 & 0 & 0\\
0 & -2 & 0 & 0\\
0 &  0 & -2 & 0\\
0 & 0 & 0 & 0
\end{pmatrix}
\end{equation}
In the interaction picture, $\hat{\cl}_{D}^I(t) = e^{-\hat{\cl}_0 t} \hat{\cl}_{D} e^{\hat{\cl}_0 t}$. The unitary part $\hat{\cl}_0$ induces oscillations with period $2\pi/\sqrt{4J^2+U^2}\ll 1/\lambda$, so we can approximate the time-dependent $\hat{\cl}_{D}^I(t)$ by its average over one period. Finally, by diagonalizing this matrix and selecting the eigenvalues with the second largest real part, we can estimate the lower bound of the decay rate for $\mean{s_z(t)}$. Specifically, we obtain
\begin{equation}
\Gamma_{Dz} = \lambda\frac{8J^2}{4J^2+U^2},\quad \Gamma_{Dx} = \lambda\frac{8J^2+U^2}{4J^2+U^2}
\end{equation}
for dephasing noise along the $z$ and $x$ direction respectively. It is easy to see that $\Gamma_{Dz}\ll \Gamma_{Dx}$ because $J\ll U$; and in the limit $J\to 0$, $\Gamma_{Dz}\to 0$ while $\Gamma_{Dx}\to \gamma$. Physically, the $s_z-$noise can only provide extra energy to stabilize the spin-flip fluctuation while the $s_x-$ can operationally flip the spin. In Fig.~\ref{figSM1}, we demonstrate all the previous arguments: dephasing noise suppresses the $\mean{s_z}$ eventually, the decay rate is proportional to the environment coupling strength, the dephasing noise along the $x-$direction has a much stronger effect than that along the $z-$direction.

\section{Numerical methods}
We remind the reader that the Hermitian part of the Linbladian is given by
\begin{equation}
H(t) = \sum_i \frac{1}{2}\left[h_x+\pi\delta(t-nT)\right]\sigma_x^i + h_y\sigma_y^i + h_z\sigma_z^i + \frac{J_{xx}}{4}\sigma_x^i\sigma_x^{i+1} + \sum_{j>i} \frac{J}{4|j-i|^\alpha}\sigma_z^i\sigma_z^j
\end{equation} 
Throughout the paper, we fix $J=1$, $h_x = 0.1$, $h_y=h_z=0.06$, $J_{xx}=0.5$ and $\alpha=1.1$. It is important to initialize the system with a highly magnetized state. This is equivalent to setting the temperature lower than the critical point, so the SSB is stable. For our simulation, at $t=0$, all the spins point up except the spin at site 6, which points down. We also use the open-boundary chain with 12 sites unless stated otherwise. We use two methods to simulate the evolution of the density matrix: exact diagonalization (ED) and stochastic trajectory (ST). Here, we are interested in a system with long-range interaction after a long-time evolution, which limits our work to smaller systems than other works. For the ED method, we trotterize the evolution as $e^{\Delta s(\hat{\cl_0}+\hat{\cl}_D)} \approx e^{\Delta s\hat{\cl}_D/2} e^{\Delta s\hat{\cl}_0} e^{\Delta s\hat{\cl}_D/2}$ where $\hat{\cl}_D$ and $\hat{\cl}_0$ are the dephasing and Hermitian parts respectively. Notably, $\cl_D$ only contains onsite terms, so its action is equivalent to the serial application of local terms. The action upon the density matrix is then given by
\begin{equation}
e^{\Delta s \cl_0}[\rho] = U(\Delta s)\rho U^\dagger(\Delta s),~U(s) = \mathcal{T}\exp\left(-i\int_0^s H(t)dt\right).
\end{equation}
We fix the basis of the density matrix in the $z-$polarization. Accordingly, the effect of $S_z$ noises is given by
\begin{equation}
e^{\Delta s \cl_{z}^n}[\rho_{ij}] = \exp\left[\frac{\lambda}{4}\left(|z^{i}_n+z^{j}_n|-1\right)\right], 
\end{equation}
where $z^{i}_n$ is the magnetization $\in\{-1/2,1/2\}$ of the $n-$spin in the $i$ basis. It is obvious to see that the diagonal terms are unaffected by the $S_z-$dephasing noise. Meanwhile, the effect of $S_x$ noises involves pairwise scrambling of the density matrix elements
\begin{equation}
e^{\Delta s \cl_{z}^n}[\rho_{ij}] = \frac{\rho_{ij}+\rho_{i'j'}}{2} +\frac{\rho_{ij}-\rho_{i'j'}}{2} e^{-\Delta s\lambda/2},
\end{equation}
where the indices $i'$ and $j'$ relate to $i$ and $j$ by the spin flip at the $n$ spin. We choose $\Delta s = T/4$ throughout the paper.

We used the ED method for up to a spin chain of 12 sites. Beyond that, it is not efficient to store the density matrix, which motivates us to employ the stochastic trajectory methods for our simulation. We use the standard stochastic jump procedure, i.e., at each time step, the wavefunction either(i) stays the same with probability $1-N\Delta s \lambda/4$, or (ii) undergoes the jump at site $n$ to become $C\ket{\psi}/|C\ket{\psi}|$ where $C=S_z^n$ or $S_x^n$ with the site index $n$ being randomly picked. The general ST procedure is described in Ref.~\cite{Jaschke2018, Daley2014}. In our work, the evolution under the Hermitian part is performed by the TDVP method \cite{Haegeman2011,itensor,Yang2020}.

\begin{figure}
	\centering
	\includegraphics[width=0.55\textwidth]{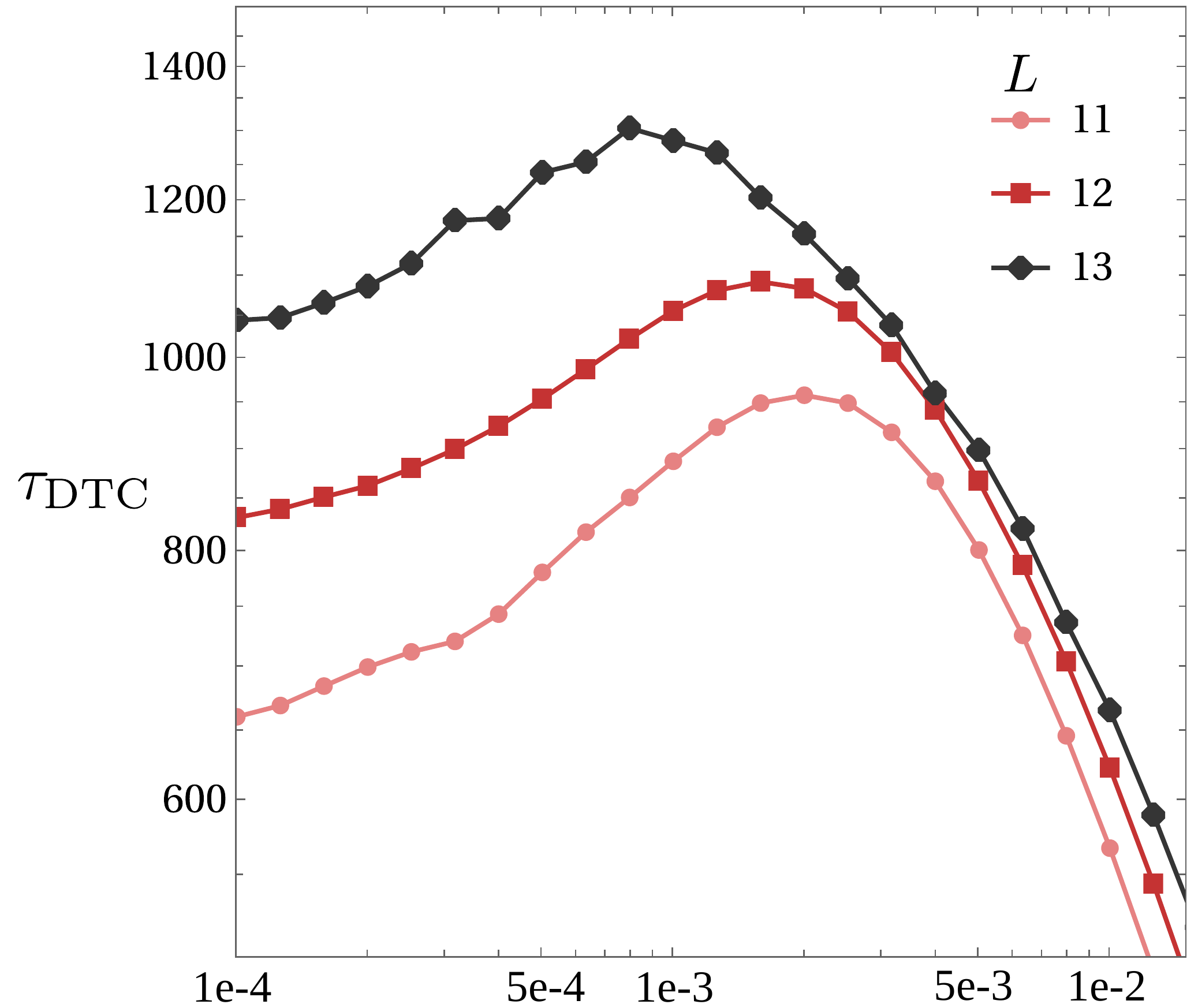}
	\caption{DTC lifetime variation with $s_z-$dephasing noise at different system sizes with the nonmonotonic behavior showing in all values of $L$\label{figSM2}}
\end{figure}

In Fig.~\ref{figSM2}, we provide the variation of $\tau_{DTC}$ as a function of $S_z-$dephasing strength at different system sizes; for $L=11$ and 12, we use the ED method, while for $L=13$ use ST method combined with TDVP to evaluate the evolution under the Hermitian part. We note that the prethermal time estimation using the Lieb-Robinson bound does not explicitly involve system size. On the other hand, $\tau_{SSB}$ might diverge with the system size, but at best by an algebraic law. Therefore, we expect the effect we report here to be observable within a wide range of system sizes, as seen in Fig.~\ref{figSM2}. We note that the shift of peak with $L$ is mostly likely because we initialize the spin chain with only one flipped spin for all values of $L$. Accordingly, the initial effective temperature is lower for larger $L$, leading to longer $\tau^0_{DTC}$ and thus the shift of the maximum toward lower $\lambda$. 

\bibliographystyle{apsrev4-2}
\bibliography{Time crystal}